# THE EFFECT OF THE NUMBER OF DISTRACTORS AND THE "NONE OF THE ABOVE" - "ALL OF THE ABOVE" OPTIONS IN MULTIPLE CHOICE QUESTIONS

A. H. Jonsdottir, TH. Jonmundsson, I.H. Armann, B.B. Gunnarsdottir, G. Stefansson

*University of Iceland, Faculty of Physical Sciences (Iceland)*

## Abstract

Multiple choice questions (MCQs) are commonly used for assessment in higher education. With increased use of on-line examination it is likely that the usage of MCQs will be even more in years to come. It is therefore of interest to examine some characteristics of these type of questions such as the effect of the number of distractors used and the "None of the above" (NOTA) or "All of the above" (AOTA) options.

The tutor-web is an open-source, on-line drilling system that is freely available to anyone having access to the Internet. The system was designed to be used for teaching mathematics and statistics but can in principle be used for other subjects as well. The system offers thousands of multiple choice questions at high school and university level. In addition to be a tool used by students for learning it has also been used as a testbed for research on web-assisted education.

The tutor-web system was used both as a learning tool and as a testing tool in a university course on mathematical statistics in the spring of 2020. Around 300 students were enrolled in the course providing tens of thousands of answers to MCQs designed to investigate the effect of the number of distractors and the use of NOTA and AOTA options in questions. The main findings of the study were that the probability of answering a question correctly was highest when a AOTA option was used as a distractor and when NOTA and AOTA were not used in questions. The probability of answering a question correctly decreased with the number of distractors.

## 1 INTRODUCTION

Increased use of new technologies, larger student groups, reduced resources and with the COVID-19 pandemic, the rise of online learning, have resulted in increased use of multiple-choice questions (MCQs) for assessment in higher education. Many experiments and analyses have been conducted designed to investigate positive and negative consequences of using MCQs for assessment (see for example [1]). The focus in this paper is more specific: to investigate the effect of the number of distractors (the number of incorrect answers) used in MCQs as well as the use of the "None of the above" (NOTA) or "All of the above" (AOTA) options. In order to do so an online drilling system, the tutor-web, was used.

### 1.1 The tutor-web

The tutor-web is an open-source, on-line drilling system that is freely available to anyone having access to the Internet. The system has been under development for over a decade in the University of Iceland and it can be accessed at http://tutor-web.net. Anyone with a valid email address can register and subsequently use the system, whose open source can be found on GitHub at https://github.com/tutor-web. Some features of the system that are worth mentioning but not discussed further here are the possibility to use real data in exercises (uploaded by the user or by linking to open data sources) and that students earn SmileyCoin, a cryptocurrency, while studying. Descriptions of these features can be found in [2] and [3].

Multiple choice questions are available for all the topics covered in the system and for some of the topics slides and tutorial notes are also available. A repeated crossover experiment was conducted over four years in order to compare learning among students handing in pen-and-paper homework with students doing homework in the tutor-web system. Significant difference in test scores was found where students scored higher after working in the tutor-web system [4].

The thousands of questions available in the system are not meant to be used for testing, rather to enhance learning, but the system has also been used as a testing tool. Considerable effort has been put into providing students with elaborated feedback after they submit their answer to an item, i.e. additional instructional information of the correct answer, as well as allocating appropriate questions to students. A more detailed description of how questions are allocated, the feedback given and other features of the system can be found in [5] and [6].

Three main approaches are used to designing drill sets for the tutor-web:

    (a) handcrafting individual items

    (b) using random numbers to generate an entire drill set based on a single item

    (c) use a generic "check the appropriate answer" header with a choice of a correct option and several distractors, where both the correct answer and distractors are chosen randomly from a reasonably large collection of possible options.

No matter which approach is chosen, varying numbers of distractors can be used in the items as well as occasional "None of the Above" or "All of the Above", which can be either correct (+) or incorrect (-) (NOTA+/NOTA-/AOTA+/AOTA-).

## 1.2 The number of distractors and the "None of the above" - "All of the above" options in multiple choice questions

Many experiments and analyses have been conducted to investigate the positive and negative consequences of using MCQs for assessment. An obvious advantage is how easy they are to score objectively but on the other hand they are difficult to make. Many aspects need to be considered when constructing a MCQ type assessment but as stated before, the focus here is on the number of distractors (incorrect answers) and the use of "None of the above" (NOTA) or "All of the above" (AOTA) options.

Research has shown an important benefit of using MCQs: they generally improve students' performance on later tests, an effect known as the *testing effect* (see for example [7], [8] and [9]). However, prior exposure to multiple-choice distractors have been shown to decrease the positive testing effect on later exams. This negative effect has been shown to increase with the number of distractors on previously seen MCQ tests [1]. This is of some concern since having more distractors on a given test decreases the probability of a guessing student to mark the correct answer by luck. Fortunately, research has also shown that providing the students with either immediate feedback or delayed feedback reduces this negative effect [10].

The use of NOTA and AOTA options in MCQs is a controversial matter. In a comprehensive and widely used multiple-choice item writing guideline [11] it is stated that "None of the Above Should Be Used Carefully" and "Avoid All of the Above". Several empirical studies have been conducted (see for example [12], [13] and [14]) mostly reporting increased item difficulty when using NOTA/AOTA options. In [15] it is however shown that the use of NOTA/AOTA options can be beneficial and strong arguments are made why they should be preferred over conventional items when accuracy in ability estimation is the goal.

## 2 METHODOLOGY

In order to investigate possible effect of the number of distractors used in MCQs as well as the use of the "None of the above" (NOTA) or "All of the above" (AOTA) options, data from an introductory university course on probability theory and statistics, taught in fall semester 2020, has been used. Assessment in the course consisted almost exclusively of tutor-web drills as no in-house finals or mid-terms were allowed due to the pandemic. The students initially were given a handful of drills as homework, but subsequently the drill sets were expanded, and new drills were generated for use in a mid-term and as a final exam. Data from students who answered less than 40 drills were excluded from the study resulting in 271 participating students.

The dataset analysed consists of answers to 4479 MCQs designed using method (c) described above. 15 different headers were used and around 300 items generated for each header type. For each header type a pool of correct answers were written as well as a pool of incorrect answers. The following is an example of a header used:

> *"Alice and Bob are about to undertake an experiment to see whether their two devices provide comparable numbers. They will collect data, which they are happy to assume all come independently from normal distributions.*
>
> *They will test an appropriate hypothesis and set up a confidence interval, in order to draw appropriate conclusions.*
>
> *Check the most appropriate answer."*

For each item a single correct answer was drawn from the pool of correct answers as well as some number of answers from the pool of distractors. The total number of distractors was chosen randomly using a truncated Poisson distribution, except in items including NOTA/AOTA options, where the NOTA/AOTA option was always the fourth and last option.

When investigating the possible effect of distractors, a subset of the dataset was used where items with NOTA/AOTA options were excluded. The total number of items, by number of distractors, can be seen in Table 1.

*Table 1. Overview of items used for distractor analysis.*

| Number of distractors | Number of items | Number of answers |
|---|---|---|
| 1 | 26 | 575 |
| 2 | 170 | 4357 |
| 3 | 279 | 6886 |
| 4 | 428 | 10568 |
| 5 | 487 | 11019 |
| 6 | 487 | 11056 |
| 7 | 763 | 18944 |

When investigating the possible effect of NOTA/AOTA options another subset of the dataset was used only including items with three distractors (four answer options in total). The total number of items by NOTA/AOTA type can be seen in Table 2 (noAOTA/NOTA are questions without NOTA/AOTA options).

*Table 2. Overview of items used for NOTA/AOTA analysis.*

| NOTA/AOTA type | Number of items | Number of answers |
|---|---|---|
| AOTA- | 597 | 14512 |
| AOTA+ | 282 | 6536 |
| NOTA- | 709 | 17009 |
| NOTA+ | 251 | 5954 |
| noAOTA/NOTA | 279 | 6886 |

As a first attempt to estimate the probability of a student answering a question correctly, by number of distractors and NOTA/AOTA type, the ratio between the number of correct answers and the total number of answers were calculated. This is a somewhat naive approach since the many thousands of answers are not independent observations but answers from 271 different students. Remember also that the questions have 15 different header types (discussed above) that might affect the difficulty of the questions. Therefore, two mixed binomial regression models were subsequently fitted, one for number of distractors (model 1) and one for NOTA/AOTA type (model 2). The response variable in both cases was a 0/1 variable indicating whether the item was answered correctly or not. In addition to the most interesting explanatory variables (number of distractors and NOTA/AOTA type), both models included a fixed effect factor variable correcting for the possible difference due to the header types and a random effect accounting for the dependency in answers from same students.

As shown in Table 2, considerable amount of data is available on students' responses to items with one up to seven distractors. In a typical class of students there will be students that know the subject and those that do not and therefore guess what the correct answer is. Using the above mentioned analysis, this data can be used to estimate the fraction of guessing students, $f_{guessing}$, using the following logic. The probability a guessing student answers a question correctly ($p_{guessing}$) can be calculated as 1/(1+number of distractors). Using the mixed binomial regression model described above the probabilities a student answers an item correctly, by number of distractors, is then estimated ($p_{est}$). With this information at hand the proportion of guessing students can be estimated by finding the value of $f_{guessing}$ that minimises the mean squared error of the following expression

$$p_{est} - \left(f_{guessing} \cdot p_{guessing} + (1 - f_{guessing}) \cdot p_{informed}\right)$$

with $p_{informed}$ being the probability a knowledgeable student answers a question correctly. With a little bit of algebra it can be seen that this minimisation is equivalent to setting up a simple linear least squares model without an intercept, with $p_{est}$ as the response variable and ($p_{guessing}$ - $p_{informed}$) as the explanatory variable.

All analyses presented in the paper were performed in the statistical environment R [16]. The lme4 package in R was used to fit the mixed logistic regression models [17].

## 3 RESULTS

### 3.1 Number of distractors

The fraction between the number of correct answers and the number of answers, by number of distractors, can be seen in the second column of Table 3. As discussed above this is a somewhat naive approach. A mixed binomial regression model with the number of distractors and header type as fixed explanatory variables and students as a random effect was therefore fitted (model 1). The difference between number of distractors was highly significant ($p < 0.001$). That was also the case for the header type ($p < 0.001$) and the student effect ($p < 0.001$). The estimated probabilities of answering a question type correctly resulting from the model are shown in the third column of Table 3.

*Table 3. Simple and model-based estimates of the probability of answering correctly.*

| Number of distractors | Proportion of correct answers | Estimated probabilities resulting from model 1 |
|---|---|---|
| 1 | 0.90 | 0.91 |
| 2 | 0.88 | 0.91 |
| 3 | 0.87 | 0.89 |
| 4 | 0.85 | 0.87 |
| 5 | 0.83 | 0.85 |
| 6 | 0.82 | 0.83 |
| 7 | 0.81 | 0.83 |

Looking at the table it can be seen that the proportions from the rather naive method and the estimated probabilities from the model are quite similar. Also, the largest difference in the estimated probabilities is between 1 and 7 distractors, around 0.08 which corresponds to a difference of 0.8 in a student grade on a 0-10 scale.

Using the logic described in the methodology chapter the fraction of guessing students can now be estimated. In this case $p_{guessing}$ is equal to (1/2, 1/3, 1/4, 1/6, 1/6, 1/7, 1/8) for 1 up to 7 distractors, $p_{est}$ is taken from the third column of Table 3 and $p_{informed}$ is equal to 1 for the seven distractor categories. The resulting fraction was found to be 17.3%.

## 3.2 "None of the above" - "All of the above" options

The ratio of the number of correct answers to the number of answers the by NOTA/AOTA type can be seen in the second column of Table 4. The predicted probability of answering a question correctly resulting from a mixed binomial regression model with the NOTA/AOTA type and header type as fixed explanatory variables and students as a random effect (model 2) are shown in the third column of Table 4. The difference between number of distractors was highly significant ($p < 0.001$). That was also the case for the header type ($p < 0.001$) and the student effect ($p < 0.001$).

*Table 4. Simple and model-based estimates of the probability of answering correctly.*

| NOTA/AOTA type | Proportion of correct answers | Estimated probabilities resulting from model 2 |
| --- | --- | --- |
| AOTA- | 0.87 | 0.88 |
| AOTA+ | 0.79 | 0.79 |
| NOTA- | 0.81 | 0.82 |
| NOTA+ | 0.85 | 0.86 |
| noAOTA/NOTA | 0.87 | 0.89 |

As before the proportions from the rather naive method and the estimated probabilities from the model are quite similar. Also, the largest difference in the estimated probabilities is between questions with no AOTA/NOTA options and questions with an AOTA+ option. The difference is around 0.1 which corresponds to a difference of 1.0 in a student grade on a 0-10 scale.

## 4 CONCLUSIONS

Multiple choice questions are commonly used for assessment in higher education and with increased use of on-line examination it is likely that the usage of MCQs will be even more in years to come. It is therefore of interest to examine some characteristics of these type of questions. The effect the number of distractors and the "None of the above" (NOTA) or "All of the above" (AOTA) options have on the probability a student answers a question correctly has been investigated in this paper. Data gathered from the tutor-web learning environment was used for this task. Data from 217 students taking a course on statistics and probability during the COVID-19 pandemic was used. The students answered 4479 MCQs in total. With this data at hand the fraction of guessing students could also be estimated.

The number of distractors (incorrect answers) were found to have a significant effect on the probability of answering correctly. The probability is highest when only one distractor is used but lowest when seven distractors are used (the highest number of distractors used in the study). The difference was found to be 0.8 in student grades on a 0-10 scale. This is not a surprising result since it is likely that a part of the student group is simply guessing the correct answer and with fewer distractors they are more likely to succeed. With all the available data it was possible to estimate the fraction of guessers in the student group to be 17.3%.

When looking at the results from the NOTA/AOTA analysis, it was found that the NOTA/AOTA type has a significant effect on the probability of answering correctly. The probability is highest when no NOTA/AOTA option is used and when an an AOTA- option is included (that is when an AOTA option is used as a distractor). Perhaps that is not surprising since it is enough for a student to be able see if one of the other distractors is incorrect to exclude the AOTA- option. The type with the lowest probability are AOTA+ items (that is when AOTA option is the correct one). The difference between no NOTA/AOTA and having an AOTA+ option was found to be 1.0 in student grades on a 0-10 scale.

The results of the analyses performed here and results from other studies indicate that including as many plausible distractors as one can in drilling systems such as the tutor-web is a good thing. Since elaborated feedback is provided to the students after they answer questions the negative effect of having many distractors as noted in [1] is not of concern. It should however be kept in mind that the distractors need to be well thought out as discussed in [18] which can be time consuming. Including AOTA/NOTA options in MCQs is a controversial thing. The results of the analysis performed here does not imply that including such options should be avoided and with the strong arguments made in [15] indicate that is it actually a good thing to include AOTA/NOTA options in drilling systems such as the tutor-web.

## ACKNOWLEDGEMENTS

A large number of individuals and institutions have made this work possible. Through the years, the projects have received funding from The Icelandic Centre for Research and from several EU H2020 grants, most recently FarFish (Horizon 2020 Framework Programme Project: 727891 — FarFish).

Continuous support has been provided by the University of Iceland where the course material has been developed, and by the University of Iceland Science Institute where most of the research and development has been conducted. The current version of the tutor-web was developed by Jamie Lentin at Shuttle Thread Ltd, who has also participated in the development of the SmileyCoin wallets.

Countless students have contributed to the tutor-web and the SmileyCoin wallet, most recently Jóhann Haraldsson who led the TA group for the intro stats course in the middle of COVID-19.